\documentclass[epsf,twocolumn,showpacs,preprintnumbers]{revtex4}

\usepackage{amsmath,amssymb,amsfonts}
\usepackage{graphics}
\usepackage{dcolumn} 
\usepackage{bm,graphicx,color}
\usepackage{epsfig,comment}
\usepackage{multirow}
\usepackage{graphicx}
\usepackage{dcolumn}
\usepackage{bm}
\usepackage{color}

\begin{document} 
\title{ How Compressed Hydrides Produce Room Temperature Superconductivity}
\author{Yundi Quan, Soham S. Ghosh, and Warren E. Pickett}
\affiliation{University of California Davis, Davis CA 95616}
\email{pickett@physics.ucdavis.edu}
\date{\today}
\begin{abstract}
The 2014-2015 prediction, discovery, and confirmation of record high temperature superconductivity above 200K in H$_3$S, followed by the 2018 extension to superconductivity in the 250-280K range in lanthanum hydride, marks a new era in the longstanding quest for room temperature superconductivity: quest achieved, at the cost of supplying 1.5-2 megabars of pressure. Predictions of numerous high temperature superconducting metal hydrides $XH_n$ ($X$=metal) have appeared, but are providing limited understanding of what drives the high transition temperature  T$_c$, or what limits T$_c$. We apply an opportunistic atomic decomposition of the coupling function to show, first, that the $X$ atom provides coupling strength as commonly calculated, but is it irrelevant for superconductivity; in fact, it is important for analysis that its contribution is neglected. Five $X$H$_n$ compounds, predicted to have T$_c$ in the 150-300K range, are analyzed consistently for their relevant properties, revealing some aspects that confront conventional wisdom. A phonon frequency -- critical temperature ($\omega_2$-T$_c$) phase diagram is obtained that reveals a common phase instability limiting T$_c$ at the {\it low pressure} range of each compound. The hydrogen scattering strength is identified and found to differ strongly over the hydrides. A quantity directly proportional to T$_c$ in these hydrides is identified.
\end{abstract}
\maketitle

\section{Background} 
The disruptive discovery of record high temperature superconductivity (HTS) in SH$_3$ above 200K\cite{Duan2014,Drozdov2015,Einaga2016,Duan2017} has now been superseded by reports from two groups of critical temperatures  T$_c$ in the 250-280K range in lanthanum hydride,\cite{ME2018a,MS2019,ME2018b} both requiring pressure in the 160-190 GPa range.   The mechanism of pairing is convincingly electron-phonon coupling (EPC),\cite{Duan2014,Ma2014,Papa2015,errea2015,jose2016,bernstein2015,akashi2015,Quan2016}, and several predictions of HTS in numerous other metal hydrides at high pressure have appeared, see for example \cite{Ma2014,Duan2017,TeHtheory,RHtheory,LaH10theory,eva1,eva2} but relatively little has been decided about the relative importance of the few underlying characteristics that determine T$_c$. This issue of analysis and understanding of the microscopic mechanisms is the topic of this paper.

For EPC superconductivity, the critical temperature T$_c$ is determined by a retarded Coulomb repulsion $\mu^{\ast}$, a minor property that varies only within the range 0.10-0.15, and the function of primary interest, the Eliashberg EPC spectral function 
$\alpha^2F(\omega) = \alpha^2(\omega)F(\omega)$, where $F(\omega)$ is the phonon density of states (pDOS) and $\alpha^2(\omega)$ gives the coupling strength from phonons of frequency $\omega$. While calculating (or measuring) $\alpha^2F$ is essential for any basic understanding of the coupling, T$_c$ itself can be obtained sufficiently accurately from the Allen-Dynes equation\cite{alldyn} T$_c$ =T$^{AD}_c(\lambda,\omega_{log},\omega_2;\mu^*)$ in terms of the EPC coupling strength $\lambda$ and two frequency moments obtained from $\alpha^2F$, the logarithmic $\omega_{log}$ and second $\omega_2$ frequency moments. Specific expressions are provided in the SM. For all aspects of EPC, the review of Giustino can be consulted.\cite{giustino2017}

\begin{figure*}[!ht]
\includegraphics[width=\textwidth]{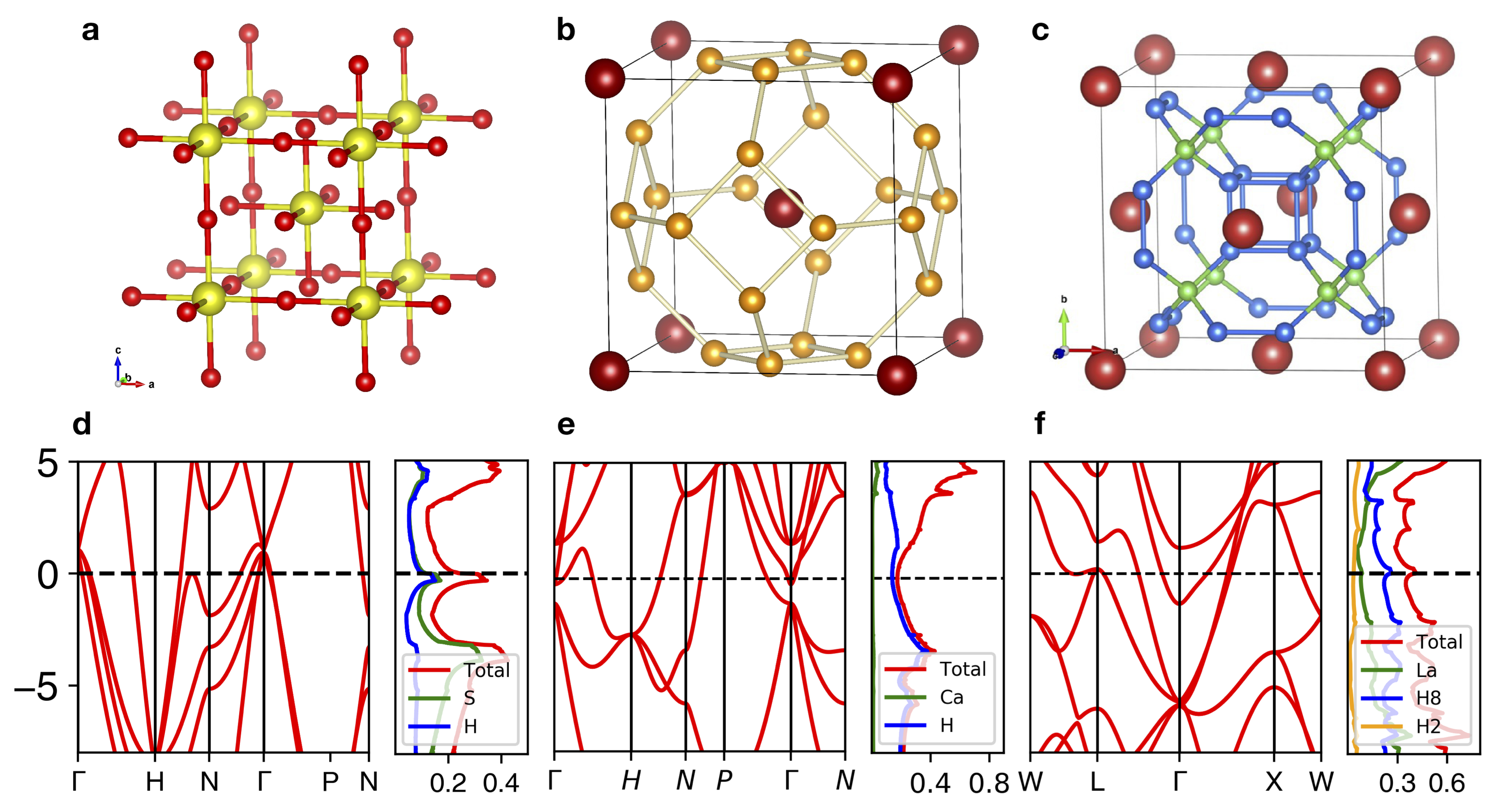}
  \caption{Top row: crystal structures of $n$=3 bcc SH$_3$, of $n$=6 bcc CaH$_6$, and $n$=10 fcc LaH$_{10}$. Lower row: corresponding band structures and electronic densities of states. In each case several bands cross the Fermi level.
  }
  \label{str}
\end{figure*}

Compounds present challenges in obtaining the relative importance of the various constituent atoms. With $\lambda$ given by
\begin{eqnarray}
\lambda=\int \frac{2}{\omega}\alpha^2F(\omega) d\omega \rightarrow \frac{N(0) I^2}{M\omega_2^2},
\label{eq:lambda}
\end{eqnarray}
individual atomic contributions are spread throughout $\alpha^2(\omega)$ and $F(\omega)$. For an elemental metal, one has the exact decomposition given at the right side of Eq.~\ref{eq:lambda} in terms of the Fermi level ($E_F=0$) density of states $N(0)$, the Fermi surface averaged squared electron-ion matrix element $I^2$, the atomic mass $M$, and the second moment $\omega_2$. The scattering strength is given by the change in crystal potential $V(\vec r)$ due to the displacement of the atom at $\vec R$
\begin{eqnarray}
I^2=\left<\left<|<k|\frac{dV}{d\vec R}|k'>|^2\right>\right>,
\end{eqnarray}
where the large brackets indicate a double average of $\vec k, \vec k'$ over the Fermi surface.

In generic compounds no such decomposition to atomic values is possible. A main point of this paper is that for binary hydrides $XH_n$,  atom-specific (subscript $j=X, H$) values 
\begin{eqnarray}
\lambda_j=N_j(0) I_j^2/M_j \omega_{2,j}^2,~~~ \lambda=\lambda_X + \lambda_H.
\label{eqn3}
\end{eqnarray} 
can be obtained and applied to great advantage to understand the origins and possible limits of T$_c$. The three crystal classes encompassing five hydrides that we have studied and compared are illustrated in Fig. 1.

\section {Atomic Analysis: H versus metal atom X}
Binary hydrides provide a unique opportunity: the light mass of H results in separation of the phonon spectrum $\omega_{q,\nu}$ and $F(\omega)$ into disjoint metal atom low frequency, and H high frequency, regimes, with examples given below. One thus obtains separate $\alpha^2F_j(\omega)$ functions for each atom type $j$=X or H from the associated frequency regime, and consequently for $\lambda = \lambda_{X} + \lambda_H$ as well. This separation provides the extension to specific atomic contributions $\lambda_j$ at the right side of Eq.~\ref{eq:lambda}. The subscript refers to each atom type $j$; $N_j(0)$ is the atom-type projected electronic DOS and the other quantities are evident. Since the denominator and $N_j(0)$ are known after calculation, we can for the first time extract the atom-specific Fermi surface averaged matrix elements $I_j^2$ for each atom type. Specifically, we obtain $I_H^2$ for each hydride for comparison. In addition, H frequency moments that are uncontaminated by $X$ contributions are obtained, for comparison across the hydrides.

We first note that from Eq.~\ref{eqn3} it seems crucial for high pressure superconductivity that $I_j^2$ {\it increases with pressure} comparably to $M_j \omega_{2,j}^2$, to maintain if not to increase  $\lambda_j$ and $\lambda$. The behavior of the atom specific $I^2$ in compounds is almost unexplored, the exception being some insight obtained from the rigid atomic potential model,\cite{GG} which has been applied successfully to close packed medium temperature (former high temperature) superconductors. While all the contributions to $I_j^2$ are available from modern EPC codes, the information has never been extracted and exploited for a deeper understanding of screening of the proton and impact on high T$_c$.

The importance of $I^2$ is evident as it is one of the three components of $\lambda_j=\eta_j/\kappa_j$: $N_j(0)$, $I_j^2$, and $M_j\omega_{2,j}^2\equiv \kappa_j$.  $\kappa_j$ is the effective harmonic lattice stiffness constant for atom $j$, thus the McMillan-Hopfield\cite{McMillan,Hopfield} parameter $\eta_j=N_j(0)I_j^2$ is an effective electronic stiffness for atom $j$, and $\lambda_j = \eta_j/\kappa_j$ is their ratio. The strong coupling limit explored by Allen and Dynes\cite{alldyn} gives $T_c\rightarrow 0.18\sqrt{\eta/M} = 0.18 \sqrt{N(0)I^2/M}$, further emphasizing the importance of $I^2$ along with $N(0)$ and $M$, also indicating the seeming {\it irrelevance of frequencies}. For these hydrides, we obtain a linear relation between T$_c$ and H (not total) parameters, discussed later.

To extract these various quantities from published papers in which only minor 
information is provided, we provide in the SM a constrained model\cite{Papa2015} 
of a piecewise constant $\alpha^2F$ that enables extraction from published figures, 
information of the type that we introduce in this paper. 

\section {Crystal structure and methods}
In Fig.~\ref{str} the bcc $Im{\bar 3}m$ space group structure of SH$_3$, the fcc $Fm{\bar 3}m$ structure of LaH$_{10}$, which has two hydrogen sites H2 (green) and H8 (blue) with two and eight sites respectively, and the bcc $X$H$_6$ structure are shown. (For structural information see the SM.) An overview of the electronic band structure and atom-projected density of states (DOS) are also shown. Several bands cross the Fermi energy (the zero of energy) so the detailed band structure {\it per se} provides little useful information about superconductivity. The LaH$_{10}$  1:10 stoichiometry is calculated to be dynamically stable in the observed pressure range and the La sublattice has been observed to be fcc,\cite{MS2019} making it the candidate structure of choice for the recent signals of superconductivity in La-H samples in the 250-280K range.\cite{ME2018a,MS2019,ME2018b} From one viewpoint, the La atom sits inside a hydrogen cage of 32 H atoms, as shown in Fig. 1, prompting the description as a clathrate structure. 

Electronic structure calculations were carried out using the pseudopotential (PP) Quantum Espresso (QE) code.\cite{QE} We have found that the results can be sensitive to the choice of PP, which partially accounts for the differing results that can be found in the literature for certain compounds. We have conservatively and consistently used Hamann's optimized norm-conserved Vanderbilt PPs identified as $oncv$.\cite{Hamann} 
The energy cutoffs for wave function and charge density expansion are 80 Ry and 480 Ry respectively. 

For self-consistent calculations, a mesh of 24$\times$24$\times$24 k points is used. The generalized gradient approximation\cite{GGA} was adopted for the exchange-correlation functional. The optimized tetrahedron method, as implemented by Kawamura {\it et al.} in QE, is used for Brillouin zone integration.\cite{kawamura} The dense mesh that we have used provides accurate energy resolutions of $N(E)$ when van Hove singularities fall at $E_F$, as occurs in both SH$_3$ and LaH$_{10}$. For phonon dispersion calculations, the 6$\times$6$\times$6 q-mesh includes the $\Gamma$ point, while to obtain electron-phonon coupling from the optimized tetrahedron method, we used a similar mesh that is displaced from $\Gamma$.  

Anharmonic corrections are known to be important for phonons and thereby T$_c$ in SH$_3$, and to stabilize it to lower pressures.\cite{errea2015} Quantum fluctuations of the H atom arise in SH$_3$\cite{akashi2015} and can shift boundaries in the phase diagrams.\cite{errea2016} In this study we restrict ourselves to the harmonic approximation and neglect quantum fluctuations; these effects shift phase diagram boundaries but do not impact our conclusions. Only with these simplifications do the formal expressions for EPC apply. 
T$_c$ is calculated consistently for all compounds from the full Allen-Dynes equation, which is a refitting to dozens of calculations to an extension of the McMillan equation for T$_c$ to include (very) strong coupling and phonon-spectrum-shape corrections. The full expression, which sports a prefactor of the logarithmic moment $\omega_{log}$ as a primary feature, is provided in the SM.

\section {New behavior holding across the hydrides} 
The compounds we discuss -- SH$_3$; CaH$_6$ and MgH$_6$; LaH$_{10}$ and YH$_{10}$  -- share broad features: they have cubic symmetry, they have a single $X$ atom per primitive cell, and many bands cross $E_F$ (see Fig.~\ref{str} for crystal structures and band structures), giving a multisheeted Fermi surface, the details of which do not seem to be important except for the possible occurrence of vHs.\cite{Quan2016} At the high pressures, lying variously across 160-400 GPa across this study, for which these structures have been reported (calculated) to be harmonically stable, the H vibrations dominate the optic modes with energies up to 220-250 meV, which are distinct from the X  dominated acoustic modes at 70 meV or lower, depending on the $X$ atom mass. Tables I-III in the Appendix contains the materials parameters obtained from our studies. The main results are as follows.

\begin{figure*}[!ht]
\centering
  \includegraphics[width=1.8\columnwidth]{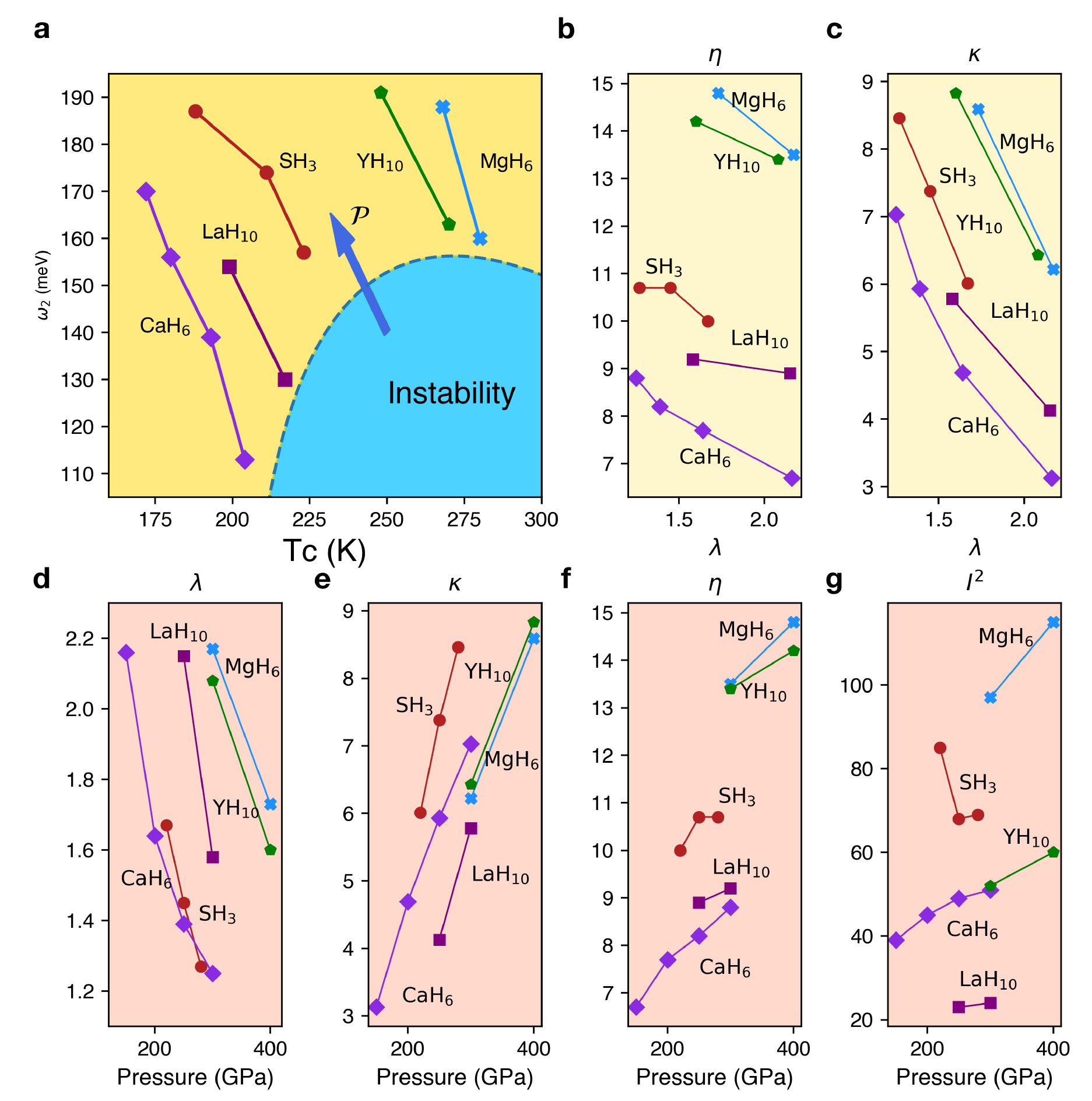}
  \caption{Interrelationships between the various materials  characteristics for hydrogen.
     (a) Schematic $\omega_2-T_c$ phase diagram, with blue indicating the island of lattice instability. The blue arrow denotes the direction of increasing pressure (P).
     (b),(c) Plots of $\kappa$ and $\eta$ respectively, versus $\lambda=\frac{\eta}{\kappa}$. Increase of $\lambda$ correlates strongly with decrease in $\kappa$ (frequencies).
     (d),(e),(f),(g) Plots of $\lambda$. $\kappa=M\omega_2^2$ (eV/\AA$^2$),
     $\eta=N(0)I^2$ (eV/\AA$^2$), and $I^2$ (eV$^2$/\AA$)^2$, respectively, versus pressure.
     All panels show each of the five hydrides toward the lower end of their region of stability.}
     \label{7panels}
\end{figure*}

\subsection{The dominance of hydrogen}
The anticipated importance of H for $T_c$ in hydrides is clouded by the observation that the $X$ atom provides 15-25\% of $\lambda$, seemingly very important.
 An overriding feature in our results of the tables in the Appendix is that coupling $\lambda_X$ from the metal atom is useless in increasing T$_c$, at best enhancing T$_c$ by only 3\% although the total $\lambda$ is increased by the above mentioned 15-25\%.  More startlingly, including the $X$ portion of $\alpha^2F$ can {\it decrease} T$_c$. For example, for LaH$_{10}$ at both 250 and 300 GPa, including $\lambda_X$ increases $\lambda$ by +14\%, but this increased strength at low frequency {\it decreases} $\omega_{log}$ by 18\% producing a net {\it decrease of T$_c$} by 5\%.   $\lambda_X$ thus becomes a source of misconceptions, and by being included in obtaining T$_c$ as in previous calculations, it has resulted in an impression (incorrect) that it contributes proportionally to T$_c$. 
 
 This anti-intuitive behavior appears to contradict the result of Bergmann and Rainer\cite{Rainer} that any small increase in coupling increases T$_c$, that is, $\delta$T$_c$/$\delta\alpha^2$F($\omega)$ is non-negative.  The resolution of this conundrum lies in effects that have been addressed before:\cite{pbamlc1972,pba1972,wep2008} in physical materials (and in a self-consistent treatment) an increase in $\alpha^2F$ at a given frequency will feed back into a softening of phonon modes. This mode softening always opposes the positive effect on T$_c$ from the increase in $\lambda$. For $X$=La in LaH$_{10}$ the softening dominates, and (as mentioned) T$_c$ {\it drops by 5\%} in spite of stronger coupling, just before the lattice instability sets in (see below). T$_c$ in CaH$_6$ and MgH$_6$ is effectively unchanged under the 15-20\% increase from $\lambda_X$; SH$_3$ shows a small positive effect. The important message is that for T$_c$, $\lambda_X$ is ineffectual and it should be disregarded to gain knowledge about increasing T$_c$. This option is included in the tables in the Appendix.

\subsection{Our major results} 
Since it was just established that $X$ atom coupling is ineffective at best and misleading in practice, henceforward we focus on the H atom contributions alone: unless otherwise stated (sometimes the H subscript is included for emphasis), our comments apply only to the H atoms' contributions (the rows in the tables in the Appendix labeled ``H''). The following observations are drawn from the $\omega_2 - T_c$ phase diagram and six other panels providing a variety of correlations in Fig.~\ref{7panels}.

\begin{figure*}[!ht]
    \centering
    \includegraphics[width=1.20\columnwidth]{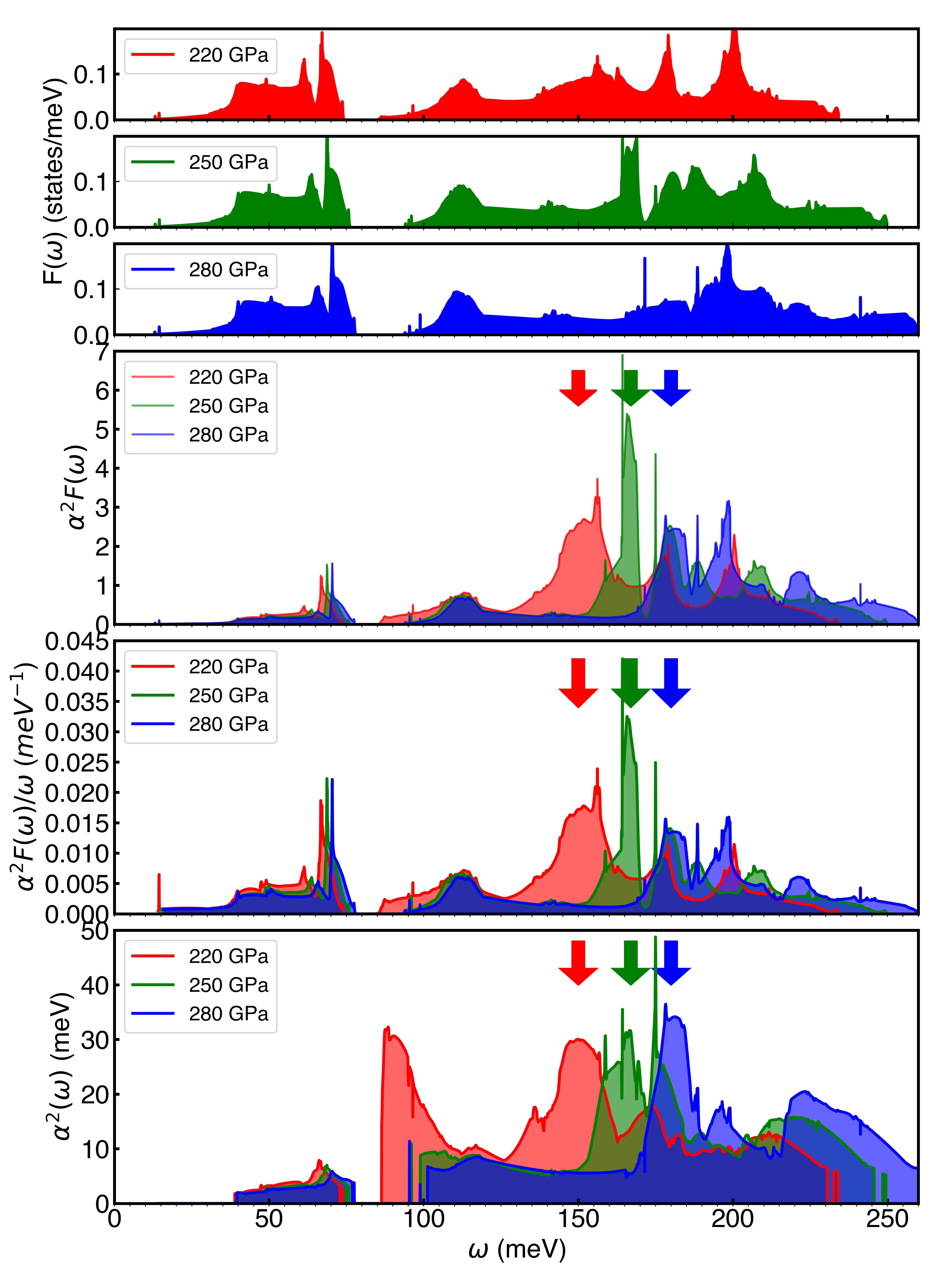}    
\caption{Views of the evolution under pressure of phonon coupling strength and frequencies for SH$_3$. From top: $F(\omega)$ for the three pressures; the Eliashberg function $\frac{\alpha^2F(\omega)}{\omega}$; 
    the ratio $\alpha^2(\omega)/\omega$ that determines $\lambda$; 
    the coupling spectrum $\alpha^2(\omega)$. In these functions there is no indication of the lattice instability just below 200 GPa.}
    \label{fig:SH3}
\end{figure*}

\subsubsection{High frequencies are important}
Higher T$_c$ compounds have higher frequency moments, as shown in Fig.~\ref{7panels}(a) and (e). Figure~\ref{7panels}(a)  provides an $\omega_2$ vs. $T_c$ phase diagram, which identifies a boundary separating the high T$_c$ region from an island of lattice instability.  Interestingly, MgH$_6$ at the highest T$_c$ end has very similar frequencies as SH$_3$. Since the denominator $M_H\omega_2^2$ in $\lambda$ is the same for these two materials, the numerator $\eta$ must be substantially larger. Figure 2(f) and the tables in the Appendix indeed show that $\eta$ is $\sim$35\% larger, with {\it twice as large matrix elements $I_H^2$} [Fig. 2(g)] overcoming a somewhat lower value of $N(0)$. This is the first clear evidence of a strong material dependence of $I_H^2$ in hydrides. 
\subsubsection{Strongest coupling involves lower frequencies}
Since pressure enhancement of hydride T$_c$ has been a prevalent notion, we quantify that T$_c$ {\it decreases with increasing pressure and increasing frequencies} within each class studied. Strong coupling is (unfortunately) associated with lower frequencies, within a region of stability. This result (noted previously in some individual materials) seems in opposition to conventional wisdom that higher pressure is better for T$_c$. Our results establish that T$_c$ is maximum at the lower pressure end of crystal stability where frequencies are softer, as shown in Fig.~\ref{7panels}(e).  T$_c$ is ultimately limited in these systems, as in many strong coupled but lower T$_c$ analogues, by lattice instability\cite{Weber1973,Testardi1975,Pickett2006} that depends on the details of EPC of the material. The emerging picture is that while pressure stabilizes favorable structures with metallic {\it atomic H}, providing high T$_c$ with high frequencies, {\it within each phase} additional pressure increases frequencies and lowers T$_c$. To repeat: the essential role of pressure is simply to stabilize structures composed of atomic H; further pressure is detrimental for T$_c$. Less pressure, that is, the instability region, comprises insulating phases with H$_2$ and H$^{-}$ units, or conducting structures with these units,\cite{eva} which do not promote strong scattering and strong EPC.   
\subsubsection{H matrix elements are not ``atomic properties''}
The derived squared H matrix element $I_H^2$ has been suggested to be an ``atomic quantity,''\cite{Hopfield,Ashcroft} not varying much from material to material. $I_H^2$ is highlighted in boldface in Tables I-III of the Appendix and plotted versus pressure in Fig.~\ref{7panels}(g), facilitating observing that it differs by a {\it factor of five} for these compounds: 24 for LaH$_{10}$ to 125 in MgH$_6$, each in eV$^2$/\AA$^2$. Evidently the screening of the proton displacement is sensitive to the response of the environment, and $I_H^2$ is not the ``atomic quantity'' as earlier suggested.
%
\subsubsection{Impact of atomic fraction of H}
Is the atomic fraction of H a crucial factor? By dividing $N_H(0)$ in the tables  in the Appendix by the number of H atoms, the contribution per H atom is obtained. The values range from around 0.022 for CaH$_6$ and MgH$_6$ to 0.033 for the vHs compounds SH$_3$, LaH$_{10}$, and YH$_{10}$; units are states/(eV atom spin). These values, which represent chemical differences and can be sensitive to the precision of the calculation and to decompositions into $X$ and H contributions, do not scale well with T$_c$.
\subsubsection{Behavior of $\lambda$(P)}
The variation of $\lambda$ with pressure depends primarily on the strong variation with pressure of the lattice stiffness $\kappa=M\omega_2^2$, see Fig.~\ref{7panels}(c). For example, $\kappa$ decreases by 55\% in CaH$_6$ from 300 to 150 GPa, beyond which the lattice becomes unstable. The minor variation of the electronic stiffness $\eta=N(0)I^2$ is apparent from Fig.~\ref{7panels} (f).
Increasing $\lambda$ by softening the lattice increases T$_c$ for currently studied hydrides but encounters lattice instability for $\lambda_H\approx$2.2.
\subsubsection{Achievement of ``atomic hydrogen''}
These alkaline earth and rare earth based compounds are effectively atomic hydrogen crystals with a charge carrier void (more precisely, a scattering strength void) in the volume consumed by the $X$ atom: the $X$ atom serves to compress and to provide bonding to hydrogen to produce atomic (versus diatomic) H, thereby producing HTS. This observation suggests that element(s) that are most able to ``break'' H$_2$ molecules into atoms in a crystalline environment provide the most promise of providing Ashcroft's ``chemical precompression concept,''\cite{Ashcroft} {\it i.e.} decreasing the pressure necessary to obtain HTS hydrides.   

\vskip 2mm \noindent

\section{Prospects for Higher T$_c$, and Limitations}
We collect here some important characteristics, by example from the various compounds.

\begin{center}
  \begin{figure}[!ht]
  \centering
   \includegraphics[width=0.97\columnwidth]{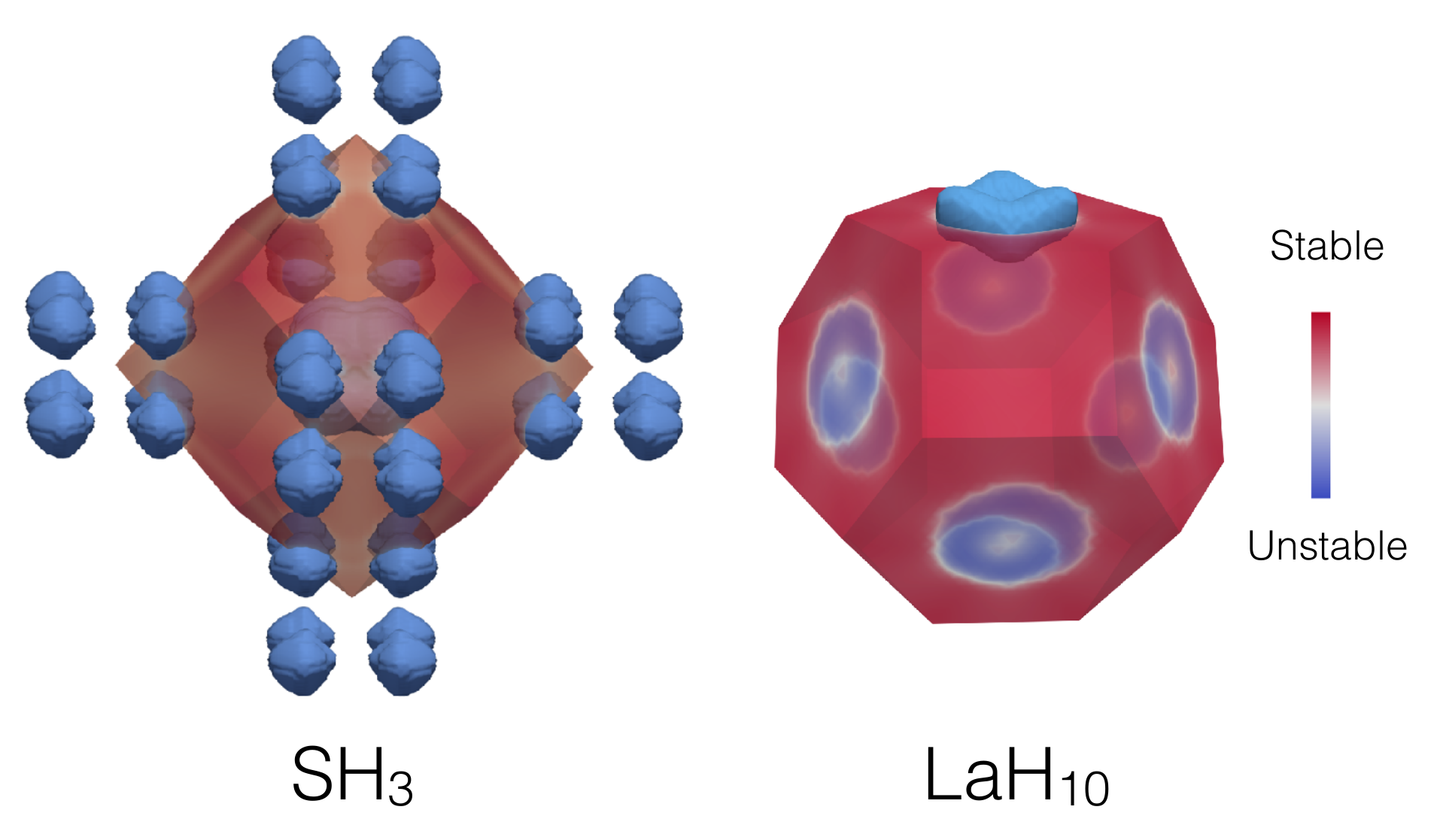}
    \caption{Regions of unstable phonons. The indicated regions of the Brillouin zone indicate where phonons first becomes unstable, in harmonic approximation. For SH$_3$ the instability regions are repeated outside the first zone for more clarity.}
  \label{instability}
\end{figure}
\end{center}

\vskip 2mm \noindent
\subsection {Strong coupling and lattice instability}
There have been many examples over several decades\cite{Weber1973,Testardi1975,Pickett2006}  where pushing a superconducting system toward stronger coupling results in marginally higher T$_c$, accompanied by renormalization toward softer phonons followed by lattice instability. The process is understood: EPC renormalizes phonon frequencies downward from their bare values $\Omega_q$:
\begin{eqnarray}
  \omega_q^2=\Omega_q^2 - 2\Omega_q \Pi_q(\omega_q)
\end{eqnarray}
where $\Pi_q(\omega)$ is the phonon self-energy that increases with $\lambda_q$: increasing coupling drives frequencies downward, as seen  from the tables in the Appendix. Then, lower frequencies increase the coupling strength measured by $\lambda$ (other things being equal): it is a cooperative process inviting vanishing frequencies and the accompanying lattice instability.

The process is illustrated for SH$_3$ in Fig.~\ref{fig:SH3}, where $F(\omega), \alpha^2F(\omega), \alpha^2F(\omega)/\omega$, and $\alpha^2(\omega)$ are shown for a range of harmonic lattice stability above the instability around 140 GPa, from which distinct features can be identified. The differences with pressure in $F(\omega)$ are unexceptional, with some hardening of the high frequency H modes proceeding as expected. Differences in $\alpha^2F$ begin to be more evident: peak values decrease from 170 to 150 to 130 meV as pressure is {\it lowered}. This shift downward of coupling strength is more striking in $\alpha^2(\omega)=\alpha^2F(\omega)/F(\omega)$, which reveals very strong coupling in the 80-120 meV region.  These H-derived optic modes are reflected in the moments of $\alpha^2F$ in Tables in the Appendix: $\omega_2$ decreases by a third before instability. Neither the moments -- {\it e.g.} $\omega_{log}$, which (over)emphasizes the low frequency modes -- nor $\lambda$ dictates the instability of the lattice by vanishing or diverging, respectively. 

Instead, a single branch (with small phase space) dips toward zero and the structure becomes dynamically unstable. In these hydrides the lower pressure, roomier structures tend to allow molecular-like dimerization of some of the H atoms into H$_2$ molecules, which is unfavorable for metallicity and strong coupling. Figure~\ref{instability} indicates the regions of the zone where instabilities in SH$_3$ and LaH$_{10}$ occur. In HS$_3$ the instability lies along the H-P symmetry line along the zone boundary, with another instability occurring at $\Gamma$. In LaH$_{10}$ the instabilities occur in a donut shape centered on the L point. In both cases as well as in CaH$_6$, the instability involves wavevectors at or near the zone boundaries, with the short wavelengths being suggestive of the instabilities being related to H$_2$ molecule type fluctuation and formation.

\vskip 2mm \noindent
\subsection {Highest T$_c$ class: rare earth decahydrides $X$H$_{10}$}
A noteworthy feature is that, for LaH$_{10}$ as in SH$_3$ which are the two materials so far observed to be superconducting nearing room temperature, the Fermi level accidentally  (if it is accidental) falls between the energies of a pair of closely spaced van Hove singularities (vHs). The associated pieces of Fermi surface in LaH$_{10}$ and resulting vHs peak in N(E) involve {\it solely} the H8 site, see Fig.~1. The additional physics\cite{Quan2016} occurring in SH$_3$ due to vHs will also apply to LaH$_{10}$ (but in less prominent form), but that is not the topic of this paper. The variation of $N(E)$ from 210 GPa to 300 GPa, is regular but minor, and the changes in Fermi surface are not visible in surface plots.  
YH$_{10}$ has a predicted T$_c$=250-270K compared to LaH$_{10}$ at 200-215K. The values of $\lambda$ at the two pressures studied are the same; the difference lies in the $\sim40$\% higher value of $\eta=N(0)I^2$ in YH$_{10}$, where a {\it much larger value of $I^2$} overcomes a somewhat lower value of $N(0)$. 
Based on current examples, increasing $\lambda$ near the instability by phonon softening does increase T$_c$ but also drives the instability, a familiar story from 1970s materials.

\vskip 2mm \noindent
\subsection {Variations within a class: alkaline earth hexahydrides $X$H$_6$}  Metal hexahydrides have been predicted to include high T$_c$ superconductors at high pressure, but synthesis and study of their properties have not yet been reported. Given the regularities discussed above, it is eye-opening to note that both the lowest and the highest T$_c$ members in Fig.~\ref{7panels} are CaH$_6$ and MgH$_6$ respectively, despite being isostructural, isovalent, and in neighboring rows in the periodic table. The difference, surprisingly, is not in higher frequencies in the smaller cell (the frequencies are similar) but in the matrix elements $I_H^2$. The origin of this difference is a topic of ongoing study for us.

\begin{figure}[!ht]
   \includegraphics[width=0.48\textwidth]{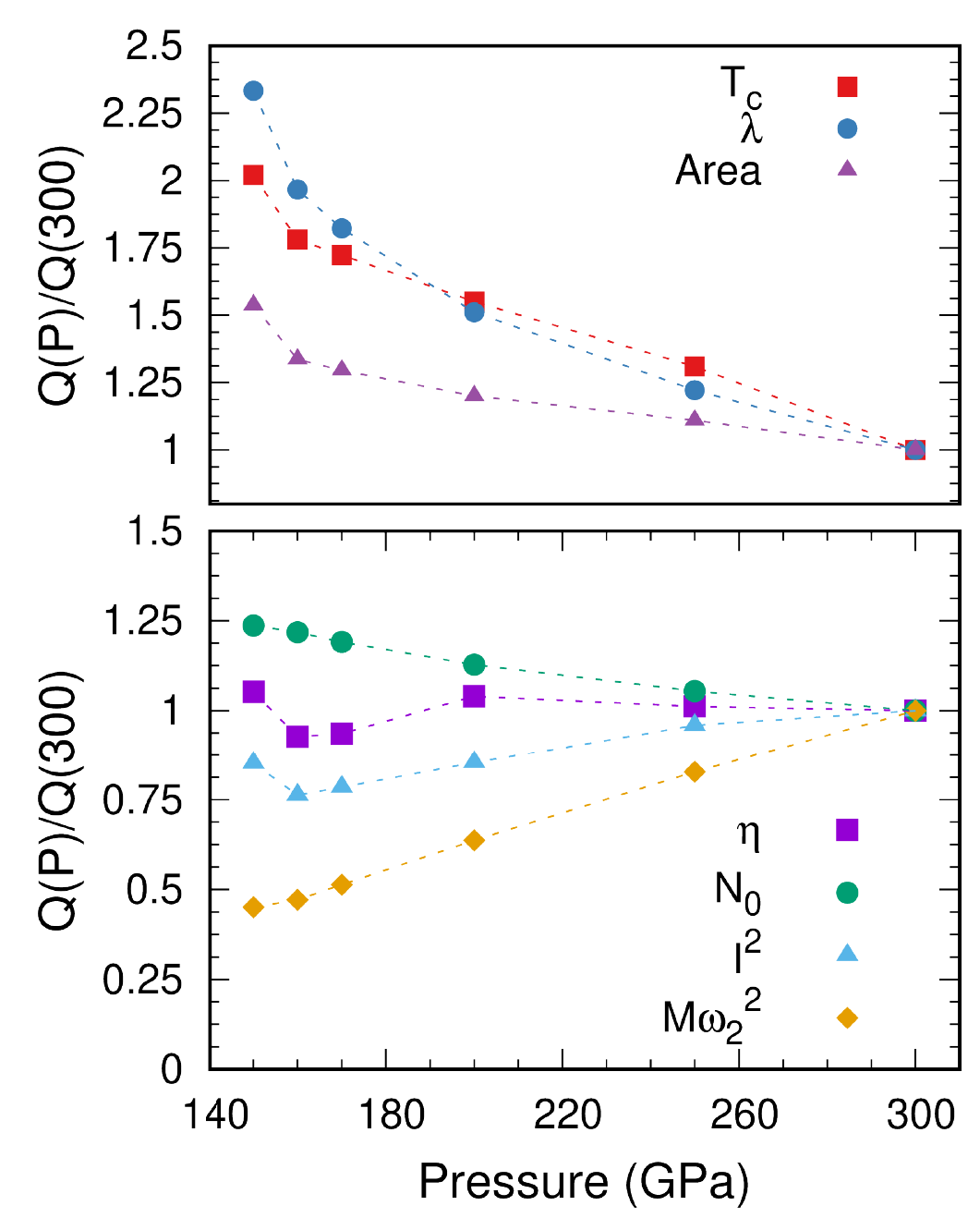}
  \caption{Pressure dependence of various superconducting quantities of cubic CaH$_6$. As emphasized in the main text, $\lambda$ and $T_c$ increase (rather strongly) at the low temperature end, before the instability. The lower panel shows that the decrease of $\omega_1^2$ is responsible, even though from frequency moments no impending instability can be inferred,  For this data, norm conserving pseudopotentials were used.}
\label{cah6}
\end{figure}

A plot of the H-related parameters for CaH$_6$ at 150-300 GPa in Fig.~\ref{cah6}, normalized to their values at 300 GPa, illuminate relative increases and decreases with pressure. The main trends follow those of SH$_3$: T$_c$ is highest at the lower pressure, with a quick upturn in $\lambda$ and T$_c$ just before the lattice becomes unstable. For this structure as for others: once the structure becomes stable,  T$_c$(P) {\it decreases} with increasing pressure, by a factor of two in our range of study. 

\vskip 2mm \noindent
\section {Areas for focus}
\subsection{$\eta_H$ versus $\kappa_H$}
The hydrides studied here reach their maximum T$_c$, just before instability, near $\lambda_H \approx$ 2.2 (somewhat smaller in SH$_3$), while T$_c$ varies from 200K to 285K. The distinction is that those with higher mean frequency just above the instability have the higher T$_c$. At first glance, the goal should be to retain strong coupling at the higher frequencies; for room temperature T$_c$ Bergmann and Rainer's analysis\cite{Rainer} suggests that for a 300K superconductor, coupling at $2\pi k_BT_c$=165 meV and above is optimal; this is in the range of the mean frequencies of the highest T$_c$ hydrides (see tables in Appendix). This choice of goal is somewhat simplistic, however, because the strongly coupled low frequency modes are in the lower frequency (not optimal frequency) range and are approaching instability precisely because they are the most strongly coupled (a chicken and egg relationship).  The best strategy seems to be to (somehow) retain strong coupling as evenly as possible over all H vibrations, preferably utilizing all momenta. Such a scenario avoids a lattice instability until a large fraction of modes become soft. 

This brings consideration to the McMillan-Hopfield constant $\eta_H=N_H(0)I_H^2$, which the analysis of Allen and Dynes indicates as the limiting behavior of T$_c$ at large coupling. Fig.~\ref{7panels}(f) shows that $\eta_H$ is much larger for the higher T$_c$ materials (LaH$_{10}$ is an exception). The next challenge therefore is to engineer $\eta_H$ because (i) so little is known about how to maximize the matrix elements $I^2$, and (ii) $N(0)$ can be sensitive to details of band structure that simply has to be calculated.  Gaining an understanding of H scattering $I_H^2$ is a current challenge but a realistic one, and one that will be crucial in learning how to retain coupling over as many H branches as possible.

\subsection{Leavens-Carbotte analysis}

\begin{figure}[!ht]
  \centering
  \includegraphics[width=1.00\columnwidth]{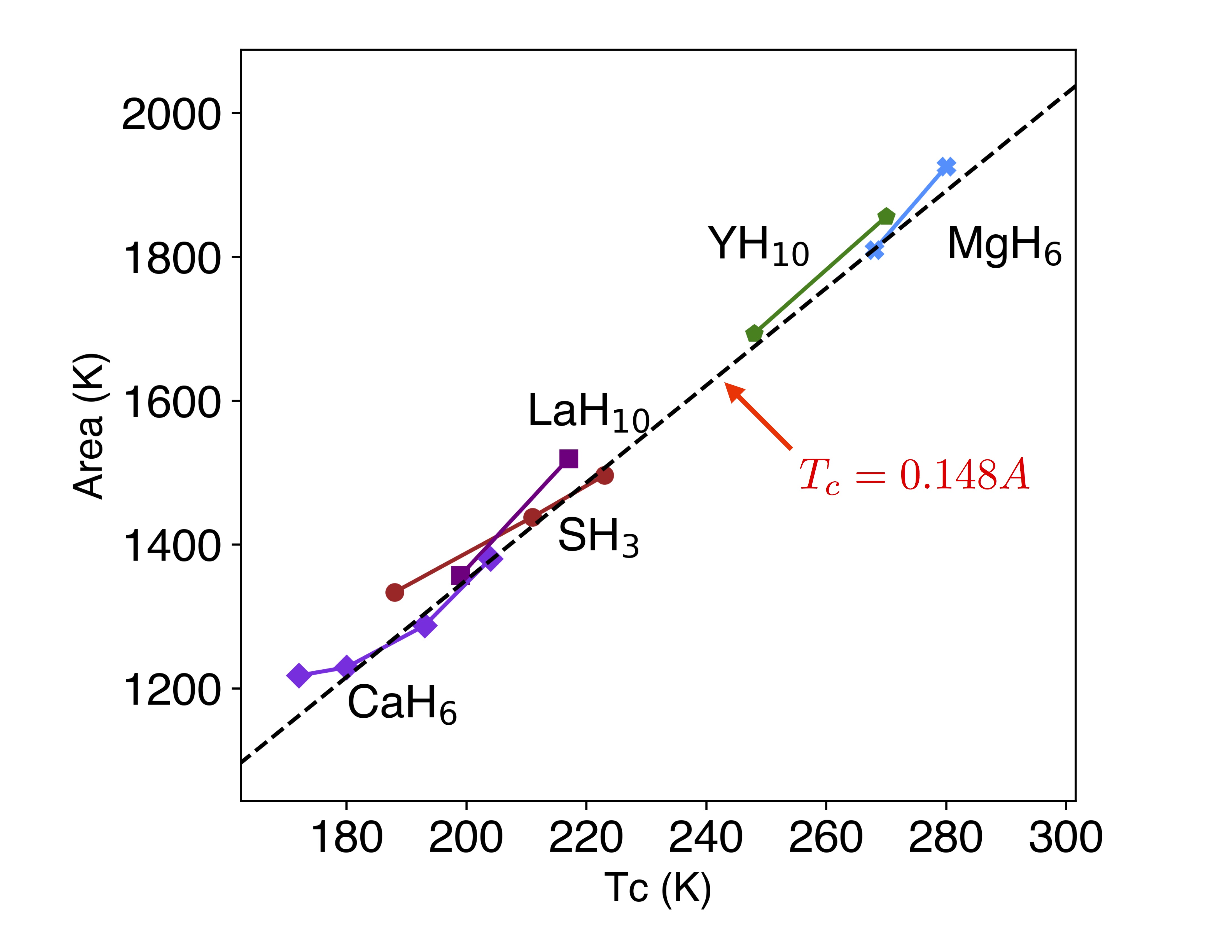}
  \caption{Plot of area under $\alpha^2$F versus T$_c$.  for binary hydrides, using H-derived quantities. The slope 0.148 denotes the Leavens-Carbotte line for strong coupled intermetallics existing in 1974.}
  \label{Leavens}
\end{figure}

An understanding of how to increase T$_c$ requires one to internalize the actual factors that determine T$_c$ at this time, which is not in the strong coupling limit. In this respect:  the somewhat involved Allen-Dynes expression is opaque -- despite its appearance, it is not exponential at all.  Leavens and Carbotte found for strong coupling materials of the time (1974)~\cite{leavens} that the area $A$ under $\alpha^2F$, which from the various definitions is $A=\lambda \omega_1/2$, was a faithful indicator of T$_c$: $T_c\approx 0.148 A$. Using our H-based values of  $A$ and T$_c$, their relationship is presented in Fig.~\ref{Leavens}, along with the Leavens and Carbotte slope of 0.148. The agreement for these five hydrides is stunningly close to their value; a least squares fit to T$_c$=S A +T$_o$ gives a practically equivalent slope of $S$=0.150 and a small intercept of T$_o$=-6K -- a direct relationship to within computational uncertainty. This relationship focuses the challenge: maximize the product $\lambda\omega_1$. Allen and Dynes proved that the strong coupling limit of Eliashberg theory is T$_c \propto \sqrt{\lambda}\omega_2$ (note: the difference between $\omega_1$ and $\omega_2$ in these hydrides is a nearly constant ratio, so for these purposes they may be considered interchangeable). Thus the strong coupling regime in hydrides has not been reached, and the Leavens-Carbotte expression provides the quantity to focus on increasing.

Our work provides another guide for reducing the pressure required for HTS hydrides. One objective is to find the element(s) $X$ in $X$H$_n$ that serves to disassociate the H$_2$ unit into atomic H in the lattice at the lowest possible pressure. Many examples indicate that a high T$_c$ phase is then likely to emerge. Our view then is that the optimum set of materials parameters, for higher T$_c$ possibly at lower pressures, is yet to be achieved.

\section{Acknowledgments}
This work used the Extreme Science and Engineering Discovery Environment (XSEDE) computational facility, which is supported by National Science Foundation grant number ACI-1548562. The project was supported by NSF grant DMR 1607139.

\appendix
\section{Structural parameters}
For each chosen pressure all lattice parameters were relaxed: first the volume and then internal
parameters.

\section{Material parameters from separation of atomic contributions}
Tables 1-3 provide the extensive numerical data calculated for the five hydrides in the three crystal structure classes that we have studied. The various rows follow the separation of the various quantities into metal atom, hydrogen (H), and total (T) compound values (the latter where appropriate). Procedures are described in the main tex and elaborated on further below.

Earlier calculations, with several differences in codes, pseudopotentials, cutoffs, and meshes have been reported. See for example:\\
\begin{itemize}
\item {\it MgH$_6$}: X. Feng, J. Zhang, G. Gao, H. Liu, and H. Wang, RSC
Adv. {\bf 5}, 59292 (2015).
\item {\it CaH$_6$}: H. Wang, J. S. Tse, K. Tanaka, T. Iitaka, and Y. Ma, Proc.
Natl. Acad. Sci. U.S.A. {\bf 109}, 6463 (2012).
\item {\it YH$_{10}$}: F. Peng, Y. Sun, C. J. Pickard, R. J. Needs, Q. Wu, and
Y. Ma, Phys. Rev. Lett. {\bf 119}, 107001 (2017).
\item {\it LaH$_{10}$}: H. Liu, I. I. Naumov, R. Hoffmann, N. W. Ashcroft, and
R. J. Hemley, Proc. Natl. Acad. Sci. U.S.A. {\bf 114}, 6990 (2017); L. Liu,
C. Wang, S. Yi, K. W. Kim, W. Kim, and J.-H. Cho,
Phys. Rev. B {\bf 99}, 140501 (2019).
\end{itemize}

\begin{table*}[!ht]
    \centering
    \begin{tabular}{|c|c|c|rrr|rrrr|rrr|}
        \hline
        &${\mathcal P}$& &$\omega_{log}$& $\omega_1$& $\omega_2$& $N_\uparrow(0)$         & $I^2$      & $\eta$   &$M\omega_2^2$  & $A$ &$\lambda$ & $T_c$\\
        & \multirow{2}{*}{GPa}   &       & \multirow{2}{*} {{\footnotesize meV}}            & \multirow{2}{*}{{\footnotesize meV}}       & \multirow{2}{*}{{\footnotesize meV}}       & \multirow{2}{*}{{\footnotesize $\frac{1}{eV}$}} & \multirow{2}{*}{{\footnotesize $\frac{eV^2}{\AA^2}$}} & \multirow{2}{*}{{\footnotesize  $\frac{eV}{\AA^2}$}} & \multirow{2}{*}{{\footnotesize $\frac{eV}{\AA^2}$}} &\multirow{2}{*} {{\footnotesize meV}} &           &\multirow{2}{*}{ {\footnotesize K}}
        \\
        & & & & & & & & & & & &
        \\
        \hline
        \hline
 \multirow{9}{*}{$SH_3$} & \multirow{3}{*}{220}
 &S &      51 &       53 &       55 &     0.12 &       85 &      9.8 &    24.0 &       11 &     0.41 &        1   \\
 &&H &     151 &      155 &      158 &     0.13 &       79 &     10.1 &     6.1 &      128 &     1.66 &      222   \\
 &&T &     122 &      135 &      144 &     0.24 &  -       &       -  &     5.0 &      140 &     2.08 &      229   \\  \cline{2-13}
 &\multirow{3}{*}{250}
 &S &      52 &       55 &       57 &     0.14 &       67 &      9.4 &    25.5 &       10 &     0.37 &        0   \\
& &H &     167 &      171 &      174 &     0.15 &       71 &     10.7 &     7.4 &      124 &     1.45 &      211   \\
& &T &     132 &      147 &      157 &     0.29 &  -       &       -  &     6.0 &      134 &     1.82 &      218   \\
   \cline{2-13}
  &\multirow{3}{*}{280}
 & S &      52 &       55 &       58 &     0.14 &       61 &      8.9 &    26.1 &        9 &     0.34 &        0   \\
 && H &     178 &      182 &      186 &     0.15 &       69 &     10.7 &     8.4 &      115 &     1.27 &      189   \\
& & T &     137 &      155 &      167 &     0.30 &  -       &       -  &     6.8 &      125 &     1.61 &      199   \\
\hline
    \end{tabular}
    \caption{Various computed properties related to superconductivity of SH$_3$, separated into contributions from the sulfur (S) and hydrogen (H) atoms separately, as well as the total. Other $X$H$_3$ compounds with the same structure have been predicted to be less promising as high temperature superconductors. $I^2$ and $\eta$ are atomic quantities, not defined for generic compounds. Certain H quantities have been emphasized in boldface font. Note the small variation in $\eta$ with pressure, and that the frequency moments scale together well.}
 \end{table*}

\begin{table*}[!ht]
 \centering
 \begin{tabular}{|c|c|c|rrr|rrrr|rrr|}
         \hline
        &${\mathcal P}$& &$\omega_{log}$& $\omega_1$& $\omega_2$& $N_\uparrow(0)$         & $I^2$      & $\eta$   &$M\omega_2^2$  & $A$ &$\lambda$ & $T_c$\\
        & \multirow{2}{*}{GPa}   &       & \multirow{2}{*} {{\footnotesize meV}}            & \multirow{2}{*}{{\footnotesize meV}}       & \multirow{2}{*}{{\footnotesize meV}}       & \multirow{2}{*}{{\footnotesize $1/eV$}} & \multirow{2}{*}{{\footnotesize $\frac{eV^2}{\AA^2}$}} & \multirow{2}{*}{{\footnotesize  $\frac{eV}{\AA^2}$}} & \multirow{2}{*}{{\footnotesize $\frac{eV}{\AA^2}$}} &\multirow{2}{*} {{\footnotesize meV}} &           &\multirow{2}{*}{ {\footnotesize K}}
        \\
        & & & & & & & & & & & &
        \\
        \hline
\multirow{12}{*}{$CaH_6$} &\multirow{3}{*}{150}
  & Ca &     31 &       32 &       33 &     0.01 & ... &      4.0 &    10.7 &        6 &     0.37 &        0   \\
 & & H &    108 &      110 &      113 &     0.17 & { \bf 39} &      6.7 &     3.1 &      119 &     2.16 &      204   \\
 & & T &     90 &       99 &      105 &     0.17 &  -       &       -  &     2.7 &      125 &     2.53 &      200   \\
 \cline{2-13}
 & \multirow{3}{*}{200}
 & Ca &      34 &       35 &       36 &     0.01 & ... &      4.2 &    13.2 &        5 &     0.32 &        0   \\
 & & H &    134 &      136 &      139 &     0.17 &  {\bf 45} &      7.7 &     4.7 &      111 &     1.64 &      193   \\
 & & T &    107 &      120 &      128 &     0.17 &  -       &       -  &     4.0 &      117 &     1.95 &      190   \\
 \cline{2-13}
 & \multirow{3}{*}{250}
   &  Ca &     37 &       38 &       39 &     0.01 &      ... &      4.7 &    15.4 &        5 &     0.30 &        0   \\
 & & H &    151 &      153 &      156 &     0.17 &  {\bf 49} &      8.2 &     5.9 &      106 &     1.39 &      180   \\
 & & T &    117 &      133 &      142 &     0.17 &  -       &       -  &     4.9 &      112 &     1.69 &      180   \\
 \cline{2-13}
 & \multirow{3}{*}{300}
& Ca &  39 &       40 &       42 &     0.01 &      ... &      5.8 &    17.1 &        6 &     0.34 &        0   \\
 & &  H &   165 &      168 &      170 &     0.13 &  {\bf 51} &      8.8 &     7.0 &      105 &     1.25 &      172   \\
 & &  T &   122 &      141 &      152 &     0.18 &  -       &       -  &     5.6 &      111 &     1.59 &      175   \\
 \hline
\multirow{6}{*}{$MgH_6$} &\multirow{3}{*}{300}
 & Mg &     48 &       49 &       50 &     0.04 &      128 &      5.6 &    15.1 &        9 &     0.37 &        0   \\
 & & H &     146 &      153 &      160 &     0.14 & {\bf 97} &     13.5 &     6.2 &      166 &     2.17 &      280   \\
 & & T &    124 &      138 &      149 &     0.18 &  -       &       -  &     5.4 &      175 &     2.54 &      279   \\
 \cline{2-13}
 & \multirow{3}{*}{400}
   & Mg &      53 &       54 &       55 &     0.04 &      121 &      4.8 &    18.1 &        7 &     0.27 &        0   \\
 & & H &      174 &      181 &      188 &     0.13 & {\bf 115} &    14.8 &     8.6 &     156 &     1.73 &      268   \\
 & & T &    149 &      164 &      176 &     0.17 &  -       &       -  &     7.5 &      163 &     2.00 &      269   \\
\hline
\end{tabular}
\caption{Contributions for CaH$_6$ and MgH$_6$ of the metal and H atoms separately (see main text), as well as the total (T) value, to the parameters determining T$_c$. The $I^2$ values for CaH$_6$ are not presented because the Ca density of states needed to derive them are too small and uncertain to obtain reliable values.}
\label{tableI}
\end{table*}

\begin{table*}[!ht]
    \centering
    \begin{tabular}{|c|c|c|rrr|rrrr|rrr|}
    \hline
            &${\mathcal P}$& &$\omega_{log}$& $\omega_1$& $\omega_2$& $N_\uparrow(0)$         & $I^2$      & $\eta$   &$M\omega_2^2$  & $A$ &$\lambda$ & $T_c$\\
        & \multirow{2}{*}{GPa}   &       & \multirow{2}{*} {{\footnotesize meV}}            & \multirow{2}{*}{{\footnotesize meV}}       & \multirow{2}{*}{{\footnotesize meV}}       & \multirow{2}{*}{{\footnotesize $\frac{1}{eV}$}} & \multirow{2}{*}{{\footnotesize $\frac{eV^2}{\AA^2}$}} & \multirow{2}{*}{{\footnotesize  $\frac{eV}{\AA^2}$}} & \multirow{2}{*}{{\footnotesize $\frac{eV}{\AA^2}$}} &\multirow{2}{*} {{\footnotesize meV}} &           &\multirow{2}{*}{ {\footnotesize K}}
        \\
        & & & & & & & & & & & &
        \\
        \hline
    \multirow{6}{*}{$LaH_{10}$} &\multirow{3}{*}{250}
      & La &      23 &       24 &       25 &     0.07 &       98 &      6.5 &    21.2 &        3 &     0.31 &        0   \\
      & & H &    112 &      121 &      130 &     0.38 & {\bf 23} &      8.9 &     4.1 &      131 &     2.15 &      217   \\
      & & T &     92 &      109 &      122 &     0.44 &  -       &       -  &     3.6 &      134 &     2.46 &      206   \\
      \cline{2-13}
      & \multirow{3}{*}{300}
      & La &      24 &       25 &       26 &     0.07 &       77 &      5.1 &    24.0 &        2 &     0.21 &        0   \\
      & &  H &   141 &      148 &      154 &     0.37 &  {\bf 24} &      9.2 &     5.8 &      117 &     1.58 &      199   \\
      & & T &  115 &      133 &      145 &     0.43 &  -       &       -  &     5.1 &      120 &     1.80 &      189   \\
    \hline
    \multirow{6}{*}{$YH_{10}$} &\multirow{3}{*}{300}
    & Y & 24 & 25 & 26 & 0.11 & 44 & 5.0 & 14.6 & 4 & 0.34 & 0   \\
    & & H & 145 & 154 & 163 & 0.25 & {\bf 52} & 13.4 & 6.4 & 160 & 2.08 &      270   \\
    & & T &    113 &      136 &      151 &     0.36 &  -       &       -  &     5.5 &      165 &     2.42 &      249   \\\cline{2-13}
    & \multirow{3}{*}{400}
    & Y  &      30 &       30 &       31 &     0.11 &       41 &      4.4 &    21.3 &        3 &     0.21 &        0   \\
    & & H  & 174 &      183 &      191 &     0.23 & {\bf 60} &     14.2 &     8.8 &      146 &     1.60 &      248   \\
    & & T & 142 &      165 &      180 &     0.34 &  -       &       -  &     7.9 &      149 &     1.81 &      236   \\
    \hline
    \end{tabular}
    \caption{Various computed properties related to superconductivity for LaH$_{10}$ and YH$_{10}$ of the metal and H atoms separately (see main text), as well as the total (T) value, to the parameters determining T$_c$.  }
\end{table*}

\end{document}